\documentclass[11pt,draftcls,onecolumn, a4paper]{IEEEtran}
\usepackage{bbm}
\usepackage{amssymb,graphicx,lineno}
\usepackage{rawfonts,amsfonts,amssymb,psfrag,color}
\usepackage{amsfonts}
\usepackage{amsmath,epsfig}
\usepackage{rawfonts,graphicx,amsfonts,amssymb}
\usepackage{bm}
\usepackage{algorithm, algorithmic}
\usepackage{url}

\begin{document}

\title{Robust Phase Retrieval via ADMM with Outliers}
\author{Xue Jiang,~\IEEEmembership{Member,~IEEE}, H. C. So,~\IEEEmembership{Fellow,~IEEE},
and Xingzhao Liu,~\IEEEmembership{Member,~IEEE}
\thanks{X. Jiang and X. Liu are with the School of Electronic Information and
Electrical Engineering, Shanghai Jiao Tong University, Shanghai, China, 200240
(E-mail: xuejiang@sjtu.edu.cn, xzhliu@sjtu.edu.cn).}
\thanks{H. C. So is with the Department of Electronic Engineering, City University
of Hong Kong, Hong Kong
(E-mail: hcso@ee.cityu.edu.hk).}}

\maketitle

\begin{abstract}
An outlier-resistance phase retrieval algorithm based on alternating direction
method of multipliers (ADMM) is devised in this letter. Instead of the widely used
least squares criterion that is only optimal for Gaussian noise environment,
we adopt the least absolute deviation criterion to enhance the robustness against outliers.
Considering both intensity- and amplitude-based observation models,
the framework of ADMM is developed to solve the resulting non-differentiable optimization problems.
It is demonstrated that the core subproblem of ADMM is the proximity operator
of the $\ell_1$-norm, which can be computed efficiently by soft-thresholding in each iteration.
Simulation results are provided to validate the accuracy and efficiency of the proposed approach
compared to the existing schemes.
\end{abstract}


\section{Introduction} \label{Sec:Intro}

In many applications, the intensity or amplitude of the signal-of-interest (SOI)
can be measured, but its phase is unavailable.
Signal reconstruction from phaseless measurements is referred to as phase retrieval \cite{Candes1,PR:SPM}, which attracts
a great attention in various fields of science and engineering, such as
optical imaging \cite{PR:SPM}, X-ray crystallography \cite{PR:xray}, astronomy \cite{PR:astronomy}, and
radar \cite{AES}.

The methods for phase retrieval in the early days consider the Fourier
transform model, i.e., the observations are
the modulus or the squared modulus of their Fourier transform.
In this case, the most well-known methods are the error reduction algorithms, including
Gerchberg and Saxton (GS) \cite{PR:GS}, Fienup \cite{PR:Fienup} which is a modified version of GS, and other variants.
The basic idea of error reduction approach is alternating projection.
It starts from a random initial estimate and iterates between the time
and frequency domains to correct the current estimate according to the time-domain \emph{a priori} knowledge and
scale Fourier coefficients to match the measured data in frequency domain.
Although the GS algorithm and its modifications are useful in practice, a theoretical issue is that the guarantee of the
convergence behavior is still unclear because the alternating projection onto nonconvex sets is involved in \cite{Candes1}.

Recently, an efficient method for phase retrieval has been developed via Wirtinger flow (WF) \cite{Candes1},
which is actually a gradient descent scheme.
More importantly, Cand\`{e}s \emph{et al.} have proved that the WF with
an initialization using the spectral method exhibits geometric convergence to the solution with a
high probability provided that the size of observations is on the order of $N\log N$
with $N$ being the length of the SOI. In order to accelerate the convergence rate of WF,
an WF with optimal stepsize (WFOS) has been devised by selecting a more appropriate
stepsize at each iteration \cite{WFOS}.

Instead of solving the nonconvex problem directly, an alternative is to relax the original problem into a convex
programming \cite{Candes:SIAM}. Note that phase retrieval requires solving a
system of quadratic equations, which can be converted into linear ones by lifting up the $N$-dimensional vector
to an $N \times N$ rank-one matrix. Therefore, it is also known as \emph{PhaseLift}.
Then the problem of rank minimization can be relaxed
into a convex trace norm minimization and solved by semidefinite programming (SDP).
Although in general the SDP-based approaches \cite{Candes:SIAM}--\cite{JSTSP} can provide satisfying results,
they become computationally demanding as the signal dimension increases because of matrix lifting.

Most existing methods are built upon the least squares (LS) criterion which is only optimal for Gaussian noise.
While they work well for the noise-free or Gaussian noise environments, their performances significantly degrade
in the presence of non-Gaussian outliers.
In fact, the phenomenon of outliers has been reported in different fields \cite{Zoubir}.
In order to enhance the robustness against outliers, we adopt the least absolute
deviation (LAD) criterion instead of the widely used
LS methodology. Considering both intensity- and amplitude-based observation models,
the framework of alternating direction method of multipliers (ADMM) is proposed to solve the resulting non-differentiable optimization
problems. The core subproblem of the proposed LAD-ADMM for phase retrieval can be converted to
the proximity operator of the $\ell_1$-norm, which can be computed efficiently by soft-thresholding in each iteration.

The remainder of this paper is organized as follows.
The problem formulation is presented in Section \ref{Sec:Model}.
Section \ref{Sec:RPR} describes the framework of LAD-ADMM for robust phase retrieval.
In Section \ref{Sec:Simulation}, simulation
results are provided to demonstrate the high accuracy of the LAD-ADMM.
Finally, conclusions are drawn in Section \ref{Sec:Conclusion}.

We use bold capital upper-case and lower-case letters to represent
matrices and vectors, respectively. The $\pmb 0$ and $\pmb I$ are the zero vector and
identity matrix, respectively. The superscripts $(\cdot)^T$ and $(\cdot)^*$ denote
the transpose and complex conjugate, respectively.
The ${\rm j}=\sqrt{-1}$ is the imaginary unit.
The $|\cdot|$ denotes the absolute value of a real number or the
modulus of a complex number and $\|\cdot\|_2$ is the Euclidean norm of a vector.
The $\odot$ is the elementwise multiplication.
Finally, $\mathbb{R}$ and $\mathbb{C}$ represent
the sets of real and complex numbers, respectively.

\section{Problem Formulation}\label{Sec:Model}

In general, there are two types of observation models for phase retrieval. One is the intensity-based model given by
\begin{equation}\label{Eq:mag}
\pmb y = |\pmb A\pmb x|^2 + \pmb v,
\end{equation}
where $\pmb x = [x_1,\cdots, x_N]^T \in \mathbb{C}^{N}$ is the SOI to be estimated,
$\pmb y = [y_1,\cdots, y_M]^T \in \mathbb{R}^{M}$ is the recorded intensity,
$\pmb A = [\pmb a_1^*,\cdots, \pmb a_M^*]^T \in \mathbb{C}^{M\times N}$
is the measurement matrix and $\pmb v$ is the noise component.
A common approach to determining $\pmb x$ from \eqref{Eq:mag} is to employ the LS criterion:
\begin{equation}\label{Eq:PR:LS}
\min_{\pmb x} \left\|\pmb y - |\pmb A\pmb x|^2\right\|_2^2.
\end{equation}
The recently reported WF algorithm \cite{Candes1},
which applies a gradient descent scheme, has been demonstrated as an efficient method for solving \eqref{Eq:PR:LS}.
The WFOS method can further accelerate the convergence rate of WF \cite{WFOS}.

Another type of observation model is based on the amplitude. That is,
\begin{equation}\label{Eq:LS:Amp}
\pmb b = |\pmb A\pmb x| + \pmb v,
\end{equation}
where $\pmb b \in \mathbb{R}^{M}$. The corresponding LS criterion is given by
\begin{equation}\label{Eq:PR:Amp:LS}
\min_{\pmb x} \left\|\pmb b- |\pmb A\pmb x|\right\|_2^2,
\end{equation}
which can be reformulated as
\begin{equation}\label{Eq:PR:ALS}
\min_{\pmb x, \pmb \phi} \left\|\pmb A\pmb x - \pmb b\odot e^{j\pmb \phi} \right\|_2^2,
\end{equation}
where $\pmb \phi = [\phi_1,\cdots, \phi_M]^T$ is phase vector.
The problem of \eqref{Eq:PR:ALS} can be solved based on alternating projections, say, GS algorithm \cite{PR:GS}.
That is, given $\pmb \phi$, $\pmb x$ can be obtained by solving the
following classic LS problem: $\min_{\pmb x} \left\|\pmb A\pmb x - \pmb b\odot e^{{\rm j}\pmb \phi} \right\|_2^2$
while $\pmb \phi = \angle (\pmb A\pmb x)$, where $\angle (\pmb A\pmb x) \in [0,2\pi)$ is the phase angle of $\pmb A\pmb x$, with
$\pmb x$ being fixed.
In the case of Fourier transform, $\pmb A$ is the Fourier matrix. The algorithm essentially
adjusts the modulus of the Fourier transform of the current
estimate so that it is consistent with the amplitude data.
Although the LS criterion of \eqref{Eq:PR:LS} and \eqref{Eq:PR:Amp:LS} works well with Gaussian noise, its performance significantly degrades with outliers.
In this letter, we aim to devise an outlier-robust phase retrieval via ADMM.

\section{Outlier-Robust Phase Retrieval} \label{Sec:RPR}

\subsection{Framework of ADMM}

To enhance the robustness against outliers, we adopt the LAD criterion instead of \eqref{Eq:PR:LS}, which is given by
\begin{equation}\label{Eq:PR:LAD}
\min_{\pmb x}\left\|\pmb y - |\pmb A\pmb x|^2\right\|_1
\end{equation}
where $\|\cdot\|_1$ is the $\ell_1$-norm of a vector defined as $\|\pmb h\|_1 = \sum_{i=1}^N |h_i|$ for $\pmb h \in \mathbb{C}^{N}$.
Since the objective function of \eqref{Eq:PR:LAD} is non-differentiable, the gradient-based approaches are not applicable anymore.
We rewrite \eqref{Eq:PR:LAD} as
\begin{equation}\label{Eq:PR:LAD1}
\begin{aligned}
&\min_{\pmb x,\pmb z}~\|\pmb z\|_1\\
&{\rm s.t.} ~\pmb z = |\pmb A\pmb x|^2 - \pmb y
\end{aligned}
\end{equation}
by introducing a real-valued auxiliary vector $\pmb z \in \mathbb{R}^{M}$. The augmented Lagrangian function
of \eqref{Eq:PR:LAD1} is \cite{Eckstein}
\begin{equation}\label{ALF}
\mathcal{L}_{\rho} (\pmb x,\pmb z,\pmb \lambda ) = \|\pmb z\|_1
+ \pmb\lambda^T\left(|\pmb A\pmb x|^2 - \pmb y - \pmb z\right) +
\frac{\rho}{2}\left\| |\pmb A\pmb x|^2 - \pmb y - \pmb z \right\|_2^2
\end{equation}
where the vector $\pmb \lambda\in\mathbb{R}^{M}$ contains the $M$
Lagrange multipliers (dual variables) and $\rho>0$ is the penalty parameter.
The augmented Lagrangian reduces to the unaugmented one with $\rho=0$.
ADMM is proved to converge for all positive values of $\rho$ under
quite mild conditions \cite{Eckstein}, which makes the selection of $\rho$
rather flexible \cite{Ghadimi}. Hence, we can simply use a fixed positive
constant for $\rho$. Of course, using possibly different penalty parameter
for each iteration may improve the convergence rate in practice \cite{Eckstein,Ghadimi}.
The Lagrange multiplier method solves \eqref{Eq:PR:LAD1} by finding a
saddle point of the augmented Lagrangian. That is,
\begin{equation}\label{Saddle}
\mathop {\max}\limits_{\pmb \lambda} \mathop {\min}\limits_{\pmb x,\pmb z}
\mathcal{L}_{\rho} (\pmb x,\pmb z,\pmb\lambda),
\end{equation}
with the following steps at each iteration:
\begin{align}\label{ADMM1}
\pmb x^{k+1} &= {\rm arg} \mathop {\min}\limits_{\pmb x}\mathcal{L}_{\rho} (\pmb x,\pmb z^k,\pmb\lambda^k)\\\label{ADMM2}
\pmb z^{k+1} &= {\rm arg} \mathop {\min}\limits_{\pmb z}\mathcal{L}_{\rho} (\pmb x^{k+1},\pmb z,\pmb\lambda^k)\\\label{ADMM3}
\pmb\lambda^{k+1} &= \pmb\lambda^k + \rho\left(|\pmb A\pmb x^{k+1}|^2-\pmb y-\pmb z^{k+1}\right),
\end{align}
where $\{\pmb x^k,\pmb z^k,\pmb\lambda^k\}$ denotes the
result at the $k$th iteration.

\subsection{Solving the Subproblems}
Now we investigate the three subproblems of \eqref{ADMM1}--\eqref{ADMM3}.
Note that the gradient of $\mathcal{L}_{\rho} (\pmb x^{k+1},\pmb z^{k+1},\pmb\lambda)$ with respect to (w.r.t.)
$\pmb \lambda$ is calculated from \eqref{ALF} as
\begin{equation}\label{lambda:grad}
\frac{\partial\mathcal{L}_{\rho} (\pmb x^{k+1},\pmb z^{k+1},\pmb\lambda)}{\partial\pmb\lambda}
=|\pmb A\pmb x^{k+1}|^2-\pmb y-\pmb z^{k+1}.
\end{equation}
Therefore, it is clear that \eqref{ADMM3} adopts a gradient ascent with a step size $\rho$ updating
the dual variable $\pmb \lambda$. The ADMM updates $\pmb x$ and $\pmb z$ in an
alternating way to avoid jointly minimizing w.r.t. two primal variables.
By ignoring the constant term independent of $\pmb x$, the subproblem of \eqref{ADMM1} can be written as
\begin{equation}\label{xk:LS}
\min\limits_{\pmb x}
\left\||\pmb A\pmb x|^2 -\left(\pmb z^k + \pmb y - \pmb \lambda^k/\rho\right)\right\|_2^2,
\end{equation}
which is equivalent to the LS problem in \eqref{Eq:PR:LS} and hence can be solved by WF or WFOS efficiently.

The subproblem of \eqref{ADMM2} can also be expressed as
\begin{equation}\label{zk:LSL1}
\min\limits_{\pmb z}
\frac{1}{2}\left\|\pmb z - \pmb c^k\right\|_2^2 + \frac{1}{\rho}\|\pmb z\|_1,
\end{equation}
where $\pmb c^k \buildrel \Delta \over = |\pmb A\pmb x^{k+1}|^2 - \pmb y + \pmb \lambda^k/\rho$.
Note that the minimizer of \eqref{zk:LSL1} defines the proximity operator of the $\ell_1$-norm, which can be computed by soft-thresholding \cite{SpaRsa}:
\begin{equation}\label{zk:prox}
z_i^{k+1} = {\rm soft}\left(c_i^k,\frac{1}{\rho}\right), ~i = 1, \cdots, M
\end{equation}
where $z_i$ and $c_i$ represent the $i$th component of $\pmb z$ and $\pmb c$, respectively, and
\begin{equation}\label{Eq:st}
{\rm soft}\left(u,a\right) \buildrel \Delta \over = {\rm sgn}(u)\cdot \max\left(|u|-a,0\right)
\end{equation}
is the well-known soft-thresholding function.

We summarize the steps of ADMM with intensity-based model for robust phase retrieval in Algorithm \ref{Algo:ADMM}.
\begin{algorithm}
    \caption{ADMM with Intensity-based Model}
    \label{Algo:ADMM}
    \algsetup{indent=2em}
     \vspace{1ex}
    \begin{algorithmic}
        \REQUIRE $\pmb A\in \mathbb{C}^{M \times N}$, $\pmb y\in \mathbb{R}^{M}$, and $\rho>0$.

    \textbf{Initialize:} $\pmb z^0=\pmb 0$, $\pmb\lambda^0=\pmb 0$.

    \textbf{for} $k=0,1,\cdots$, \textbf{do} until converge
    \begin{enumerate}
    \item $\pmb x^{k+1} = \arg\min\limits_{\pmb x}
\left\||\pmb A\pmb x|^2 -\left(\pmb z^k + \pmb y - \frac{\pmb\lambda^k}{\rho}\right)\right\|_2^2$, which is solved by WF or WFOS.
\vspace{1ex}
    \item Calculate $\pmb c^k = |\pmb A\pmb x^{k+1}|^2 - \pmb y + \pmb \lambda^k/\rho$.
    \vspace{1ex}
    \item $z_i^{k+1} = {\rm soft}\left(c_i^k,\frac{1}{\rho}\right), ~i = 1, \cdots, M$.
    \vspace{1ex}
    \item $\pmb\lambda^{k+1} \leftarrow \pmb\lambda^k + \rho\left(|\pmb A\pmb x^{k+1}|^2-\pmb y-\pmb z^{k+1}\right)$.
    \end{enumerate}
    \textbf{end for}
\ENSURE $\pmb x^{k+1}$.
    \end{algorithmic}
\end{algorithm}

\subsection{Extension to Amplitude-based Model}

Similarly, we adopt the LAD criterion for the amplitude-based model:
\begin{equation}\label{Eq:PR:LAD1:Amp}
\begin{aligned}
&\min_{\pmb x,\pmb z}~\|\pmb z\|_1\\
&{\rm s.t.} ~\pmb z = |\pmb A\pmb x| - \pmb b
\end{aligned}
\end{equation}
and the corresponding augmented Lagrangian function is
\begin{equation}\label{ALF:Amp}
\bar {\mathcal{L}}_{\rho} (\pmb x,\pmb z,\pmb \lambda ) = \|\pmb z\|_1
+ \pmb\lambda^T\left(|\pmb A\pmb x| - \pmb b - \pmb z\right) +
\frac{\rho}{2}\| |\pmb A\pmb x| - \pmb b - \pmb z \|_2^2.
\end{equation}
To find the saddle point of
\begin{equation}\label{Saddle1}
\mathop {\max}\limits_{\pmb \lambda} \mathop {\min}\limits_{\pmb x,\pmb z}
\bar {\mathcal{L}}_{\rho} (\pmb x,\pmb z,\pmb\lambda),
\end{equation}
the following steps are performed iteratively:
\begin{align}\label{ADMM4}
\pmb x^{k+1} &= {\rm arg} \mathop {\min}\limits_{\pmb x}\bar {\mathcal{L}}_{\rho} (\pmb x,\pmb z^k,\pmb\lambda^k)\\\label{ADMM5}
\pmb z^{k+1} &= {\rm arg} \mathop {\min}\limits_{\pmb z}\bar {\mathcal{L}}_{\rho} (\pmb x^{k+1},\pmb z,\pmb\lambda^k)\\\label{ADMM6}
\pmb\lambda^{k+1} &= \pmb\lambda^k + \rho\left(|\pmb A\pmb x^{k+1}|-\pmb b-\pmb z^{k+1}\right).
\end{align}
It is not difficult to verify that the subproblem of \eqref{ADMM4} is equivalent to
\begin{equation}\label{xk:LS}
\min\limits_{\pmb x}
\left\||\pmb A\pmb x| -\left(\pmb z^k + \pmb y - \pmb \lambda^k/\rho\right)\right\|_2^2,
\end{equation}
which can be solved in the same way as \eqref{Eq:PR:Amp:LS} by alternating minimization,
while \eqref{ADMM5} is computed via soft-thresholding, which is given by
\begin{equation}\label{}
z_i^{k+1} = {\rm soft}\left(|\pmb A\pmb x^{k+1}| - \pmb b + \pmb \lambda^k/\rho, 1/\rho\right), ~i = 1, \cdots, M.
\end{equation}
Now we summarize the steps of ADMM with amplitude-based model in Algorithm \ref{Algo:ADMM:AMP}.
\begin{algorithm}
    \caption{ADMM with Amplitude-based Model}
    \label{Algo:ADMM:AMP}
    \algsetup{indent=2em}
     \vspace{1ex}
    \begin{algorithmic}
        \REQUIRE $\pmb A\in \mathbb{C}^{M \times N}$, $\pmb b\in \mathbb{R}^{M}$, and $\rho>0$.

    \textbf{Initialize:} $\pmb z^0=\pmb 0$, $\pmb\lambda^0=\pmb 0$.

    \textbf{for} $k=0,1,\cdots$, \textbf{do} until converge
    \begin{enumerate}
    \item $\pmb x^{k+1} = \arg\min\limits_{\pmb x}
\left\||\pmb A\pmb x| -\left(\pmb z^k + \pmb b - \frac{\pmb\lambda^k}{\rho}\right)\right\|_2^2$, which is solved by GS.
\vspace{1ex}
    \item Calculate $\pmb c^k = |\pmb A\pmb x^{k+1}| - \pmb b + \pmb \lambda^k/\rho$.
    \vspace{1ex}
    \item $z_i^{k+1} = {\rm soft}\left(c_i^k,\frac{1}{\rho}\right), ~i = 1, \cdots, M$.
    \vspace{1ex}
    \item $\pmb\lambda^{k+1} \leftarrow \pmb\lambda^k + \rho\left(|\pmb A\pmb x^{k+1}|-\pmb b-\pmb z^{k+1}\right)$.
    \end{enumerate}
    \textbf{end for}
\ENSURE $\pmb x^{k+1}$.
    \end{algorithmic}
\end{algorithm}

\section{Simulation Results}\label{Sec:Simulation}

In this section, we compare the proposed LAD-ADMM with WF and GS under the intensity- and amplitude-based observation models, respectively.
The normalized mean square error (NMSE) is used as the performance metric, which is defined as
\begin{equation}
{\rm NMSE} = \frac{\|\pmb x^k-\pmb x\|^2}{\|\pmb x\|^2}
\end{equation}
where $\pmb x$ is the true SOI. The Gaussian mixture model (GMM) is taken as the impulsive noise.
The probability density function of the two-term GMM is given by
\begin{equation}
p_v(v) = \sum\limits_{i=1}^2\frac{c_i}{\sqrt{2\pi}\sigma_i}\exp\left(-\frac{v^2}{2\sigma_i^2}\right),
\end{equation}
where $c_i\in [0,1]$ and $\sigma_i^2$ are the probability and variance of the $i$th term, respectively. We have $c_1+c_2=1$.
If $\sigma_2^2\gg\sigma_1^2$ and $c_2<c_1$, noise samples of larger variance $\sigma_2^2$ occurring
with a smaller probability $c_2$ can be considered as outliers embedded in Gaussian background noise of variance
$\sigma_1^2$. Hence the GMM is widely used to model the scenario with both Gaussian noise and
outliers.
The total noise variance is $\sigma_v^2=\sum_i c_i\sigma_i^2$ and the signal-to-noise ratio (SNR)
is defined as $\|\pmb x\|_2^2/\sigma_v^2$. In the simulations, we
set $\sigma_2^2=100\sigma_1^2$ and $c_2=0.1$. Therefore, 10\% of noise samples can be considered as outliers.
The measurement vector $\pmb a_i$ consists
of independent standard complex Gaussian variables, i.e.,
$\pmb a_i \in \mathbb{C}^{N} \sim \mathcal{N}(\pmb 0,\pmb I/2) + j\mathcal{N}(\pmb 0,\pmb I/2)$, $i = 1,\cdots, M$.
The length of the signal is $N = 32$. The number of observations is
eight times the signal dimension, i.e, $M = 8N$. We set $\rho = 1$.
For fair comparison, the same initial value obtained from the
spectral method \cite{Candes1} is taken for different approaches.
When plotting the performance curves, 100 Monte Carlo trials are performed.

\begin{figure}
\begin{center}
\includegraphics[width=9cm]{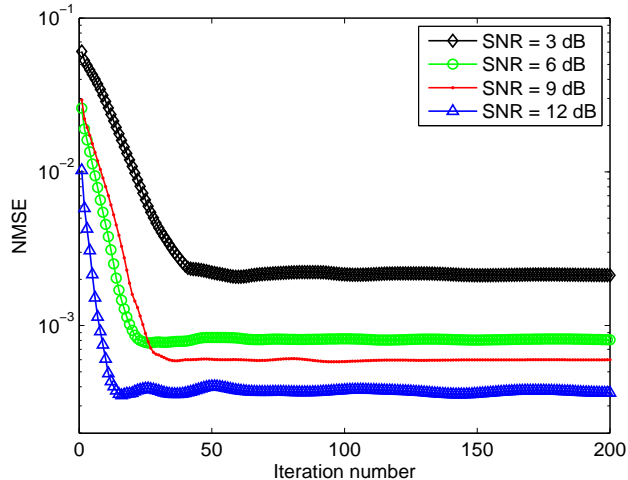}
\caption{NMSE versus iteration number with intensity-based model.}\label{Fig:NMSE:Iter}
\end{center}
\end{figure}

\begin{figure}
\begin{center}
\includegraphics[width=9cm]{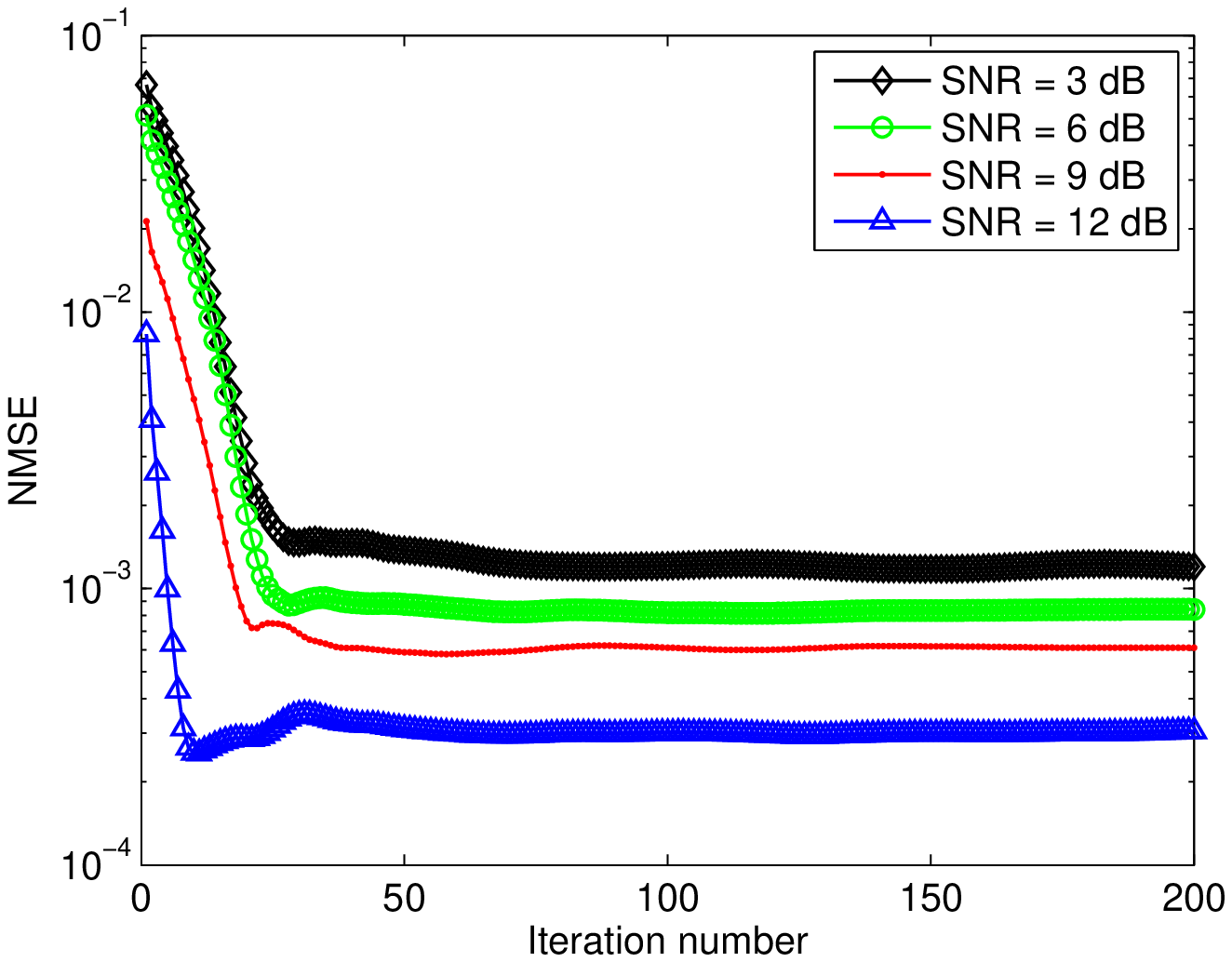}
\caption{NMSE versus SNR with amplitude-based model.}\label{Fig:NMSE:Iter:Mag}
\end{center}
\end{figure}

\begin{figure}
\begin{center}
\includegraphics[width=9cm]{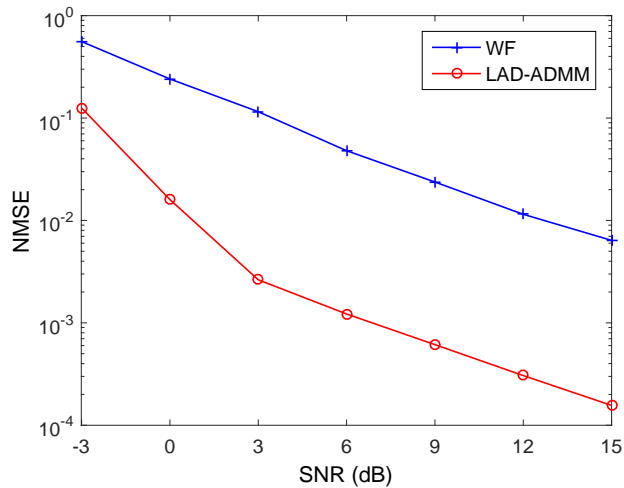}
\caption{NMSE versus SNR with intensity-based model.}\label{Fig:NMSE:SNR}
\end{center}
\end{figure}

\begin{figure}
\begin{center}
\includegraphics[width=9cm]{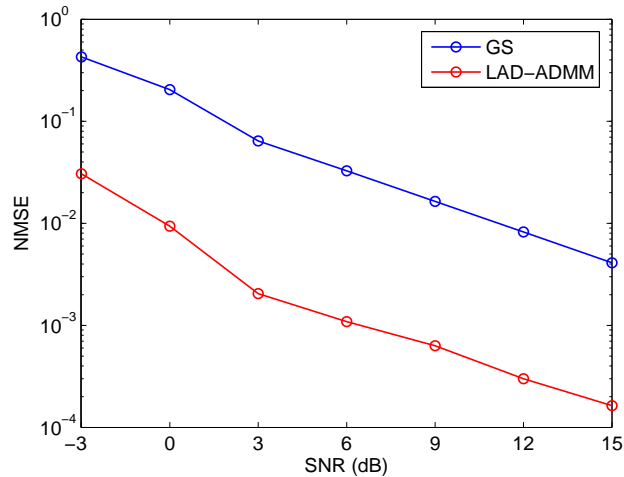}
\caption{NMSE versus SNR with amplitude-based model.}\label{Fig:NMSE:SNR:Mag}
\end{center}
\end{figure}

First, we investigate the convergence behavior of the LAD-ADMM.
In Figs. \ref{Fig:NMSE:Iter} and \ref{Fig:NMSE:Iter:Mag}, we plot the NMSE versus iteration number
with intensity- and amplitude-based observation models, respectively.
We can observe that the LAD-ADMM converges in several tens of iterations and the NMSE reaches a lower bound
in tens of iterations at ${\rm SNR}=12$~dB.
Figs. \ref{Fig:NMSE:SNR} and \ref{Fig:NMSE:SNR:Mag} show the NMSE versus SNR with intensity-
and amplitude measurements, respectively.
It can be seen that the LAD-ADMM
outperforms the WF and GS, which achieves a more accurate solution for different SNR conditions.
As shown in Fig. \ref{Fig:NMSE:SNR}, the NMSE of LAD-ADMM is about the order of $10^{-4}$ at ${\rm SNR}=15$~dB
while that of WF is $10^{-2}$.


\section{Conclusion}\label{Sec:Conclusion}
An outlier-robust phase retrieval algorithm based on the ADMM is devised in this letter.
Instead of the widely used LS criterion that is only optimal for Gaussian noise environment,
we adopt the LAD criterion to enhance the robustness against outliers.
The framework of ADMM is developed to solve the resulting non-differentiable optimization problems which is applicable for
both intensity- and amplitude-based observation models.
We have demonstrated that the subproblems of LAD-ADMM can be computed efficiently in each iteration.
Simulation results validated the convergence behavior of the proposed algorithm and
its accuracy compared to the existing techniques.

\end{document}